\documentclass[structabstract]{aa}  
\usepackage{graphicx}
\usepackage{txfonts}
\usepackage{natbib}
\usepackage{amssymb}
\usepackage[usenames]{color}
\begin{document}
   \title{Disk evaporation in a planetary nebula
      \thanks{Based on observations collected at the European Organisation for 
Astronomical Research in the Southern Hemisphere, Chile
        (proposals 075.D-0104, 077.D-0652, 081.D-0130) and HST (program 9356)} }

   \subtitle{}

   \author{K. Gesicki\inst{1} \and A. A. Zijlstra\inst{2}
           \and C. Szyszka\inst{3,2} \and M. Hajduk\inst{1,4}
           \and E. Lagadec\inst{3} \and L. Guzman Ramirez\inst{2}
	  }


   \institute{Centrum Astronomii UMK, 
              ul.Gagarina 11, 
	      PL-87-100 Torun, 
	      Poland
              \\
              \email{ 
	      Krzysztof.Gesicki@astri.uni.torun.pl
	      }
              \and
              Jodrell Bank Centre for Astrophysics,
              School of Physics \&\ Astronomy,
	      University of Manchester,
              Oxford Road,
              Manchester M13\,9PL, UK \\
              \email{a.zijlstra@manchester.ac.uk}
             \and
              European Southern Observatory, Karl Schwarzschildstrasse 2,
              Garching 85748, Germany
              \and
              Nicolaus Copernicus Astronomical Center, 
              ul. Rabia\'{n}ska 8, 87-100 Torun, Poland  
           }


  \abstract
   {}
   {Binary interactions are believed to be important contributors to the
     structures seen in planetary nebulae (PN), and the sole cause of the newly
     discovered compact dust disks. The evolution of such disks is not clear,
     nor are the binary parameters required for their creation.}
   {We study the Galactic bulge planetary nebula M\,2-29 (for which a 
   3-year eclipse event of the central star has been attributed to a dust disk) using
   HST imaging and VLT spectroscopy, both long-slit and integral field. }
   { The central PN cavity of M\,2-29 is filled with a decreasing, slow wind. An inner high density core is detected, with radius less than 250\,AU, interpreted as a rotating gas/dust disk with a bipolar disk wind. The evaporating disk is
    argued to be the source of the slow wind. The central star is a source of a very fast wind ($\sim 10^3\,\rm km\,s^{-1}$). An outer, partial ring is seen
    in the equatorial plane, expanding at 12\,km\,s$^{-1}$. The azimuthal
    asymmetry is attributed to mass-loss modulation by an eccentric
    binary. M\,2-29 presents a crucial point in disk evolution, where
    ionization causes the gas to be lost, leaving a low-mass dust disk
    behind. }
    {}

   \maketitle
%

\section{Introduction}

Disks are a recent addition to the range of structures seen in planetary
nebulae. The outer nebulae show a large variety of morphological
characteristics, including tori, bipolar flows, and ellipsoidal shapes
\citep[e.g.][]{Ramos2008}, with typical sizes of $10^4$--$10^5$ AU. But
embedded in these nebulae, in a number of cases a central dust disk has been
discovered with sizes of order $10^2$ AU, which can be resolved with the VLT
interferometer \citep[e.g.][]{Chesneau2007}. The disks appear well aligned with
the outer nebulae but are too low mass to directly have affected their
formation. More likely they are a byproduct of the original shaping process.

Other evidence exists for the presence of material very close to some central
stars, including unresolved emission-line cores \citep{Rodriguez2001} and
eclipses from dust disks \citep{Hajduk2008}. The disks are believed to require
binary systems for their formation but the details of the required system
parameters are unclear. An interesting review on binary systems and their
shaping of PNe was recently published by \citet{deMarco2009}.

Here we present a study of a planetary nebula with such an unresolved
emission-line core, known to harbour a dust disk: M\,2-29 (PN\,G\,004.0$-$03.0).
Deep echelle spectra and integral-field spectra, together with HST imaging,
allow us to disentangle the various components of this complex nebula, and to
determine the velocity fields. We find the central core to be the source of an
on-going, but decreasing wind, which we attribute to a disk wind. Although
direct evidence for the binary companion remains elusive, the outer structure
suggests an interaction with a star on a non-circular orbit. From comparison
with other objects with dust disks, we find evidence that such disks strongly
decrease in mass during the planetary nebula evolution. However, they
remain sufficiently massive that residual disks can survive long into the
white dwarf phase.

\section{Observations}

\subsection{The HST images}

\begin{figure*}
\resizebox{\hsize}{!}{
\includegraphics{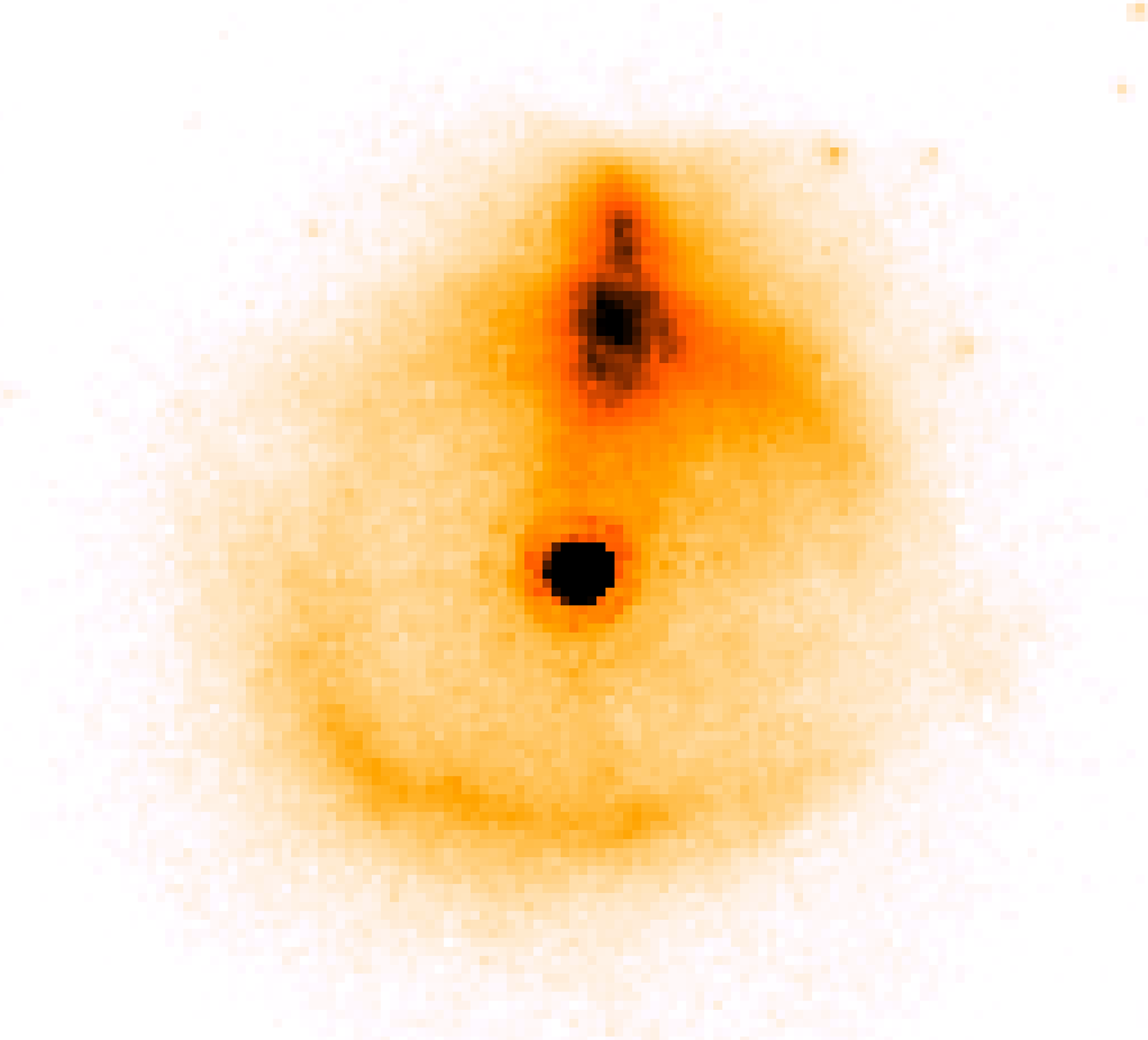} 
\includegraphics{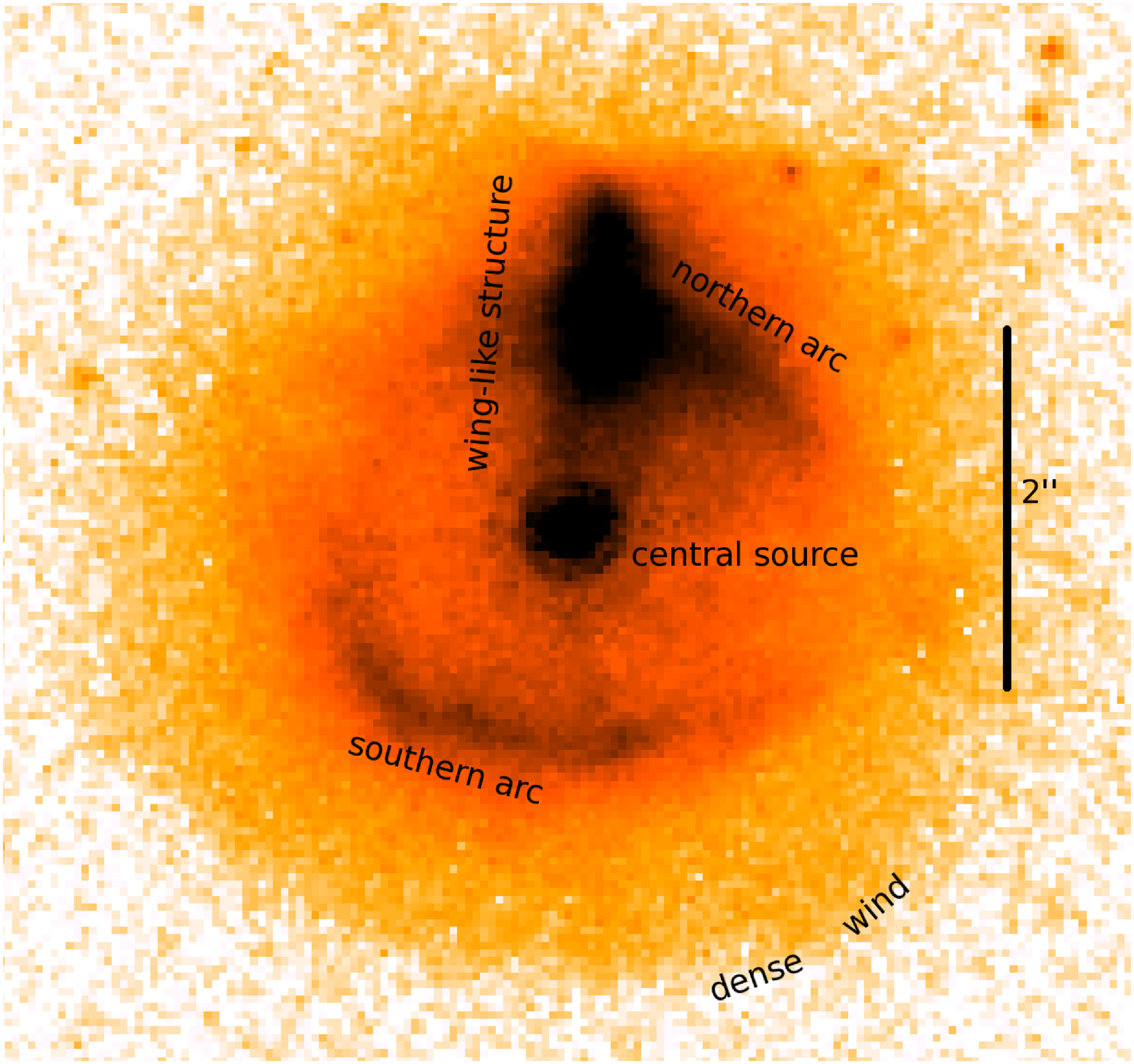}
}
      \caption{H$\alpha$ HST image of M\,2-29 obtained in 2003. Left shows a
        linear intensity scale, right shows a logarithmic intensity scale to
        bring out the fainter emission.  North is on top, east to the
        left. The length of the bar is 2\,arcsec. Taken from
        \citet{Hajduk2008} }
    \label{Halpha_image}
\end{figure*}

SNAPshot HST/WFPC2 images of M\,2-29 were obtained in 2003, in three
different filters: F656N, F547M and F502N. The pixel scale of the camera is
0.0455\,arcsec.  The H$\alpha$ image \citep{Hajduk2008} is show in
Fig.\,\ref{Halpha_image}. The main nebular components are the central source,
a wing-like structure, and the extended, fainter nebula.

\subsection{VLT/UVES long-slit spectra}

A 600-second VLT/UVES \citep{Dekker2000} echelle spectrum was obtained in 2005,
with the slit 0.5\,arcsec wide and 11\,arcsec long. The slit was put through
the centre of the nebula, aligned with the jet-like structure. The pipeline
calibration was applied, including wavelength calibration and merging of the
separate orders of the spectrum. The observations cover a spectral region
approximately from 3300 to 6600\,\AA.  The long-slit spectra of M\,2-29 are
spatially resolved although they have much worse resolution than HST images
and the seeing during the observations was nearly 1.5 arcsec.

\subsection{VLT/ARGUS-IFU spatially resolved spectroscopy}

We also retrieved publicly available data from the ARGUS Integral Field Unit
(IFU), a mode of the FLAMES instrument at the VLT \citep{Pasquini2002}. These
observations of M\,2-29 were performed in visitor mode as a part of program
077.D-0652 (PI. D.\,Sch\"onberner). Two spectral ranges are available, HR8 and
HR14B covering range 4911--5158 and 6383--6626\,\AA\ with resolution
R$\sim$32000 and R$\sim$46000 respectively. Because we aimed for small scale
structure, we selected only those frames which had seeing better than
1\,arcsec; this constraint yielded one frame in HR14 (700s) and six frames in
HR8 (6x110s). The large-area (12x7 arcsec$^2$) IFU was used, where each single
resolving element (spaxel) covers 0.52x0.52\,arcsec$^2$.

The data were analyzed with the girBLDRS pipeline (a.k.a. Geneva pipeline).
Correction of bias level flat fielding and the wavelength calibration were
prepared in a standard way. Spectra were extracted with the sum-extraction
method, because the default optimum extraction is not well suited to
emission-like object. The airmass was lower than 1.06 for all frames
alleviating the need to correct for differential atmospheric refraction.  The
data were not flux calibrated.

\subsection{VLT/VISIR images}

\begin{figure}
\resizebox{\hsize}{!}{\includegraphics{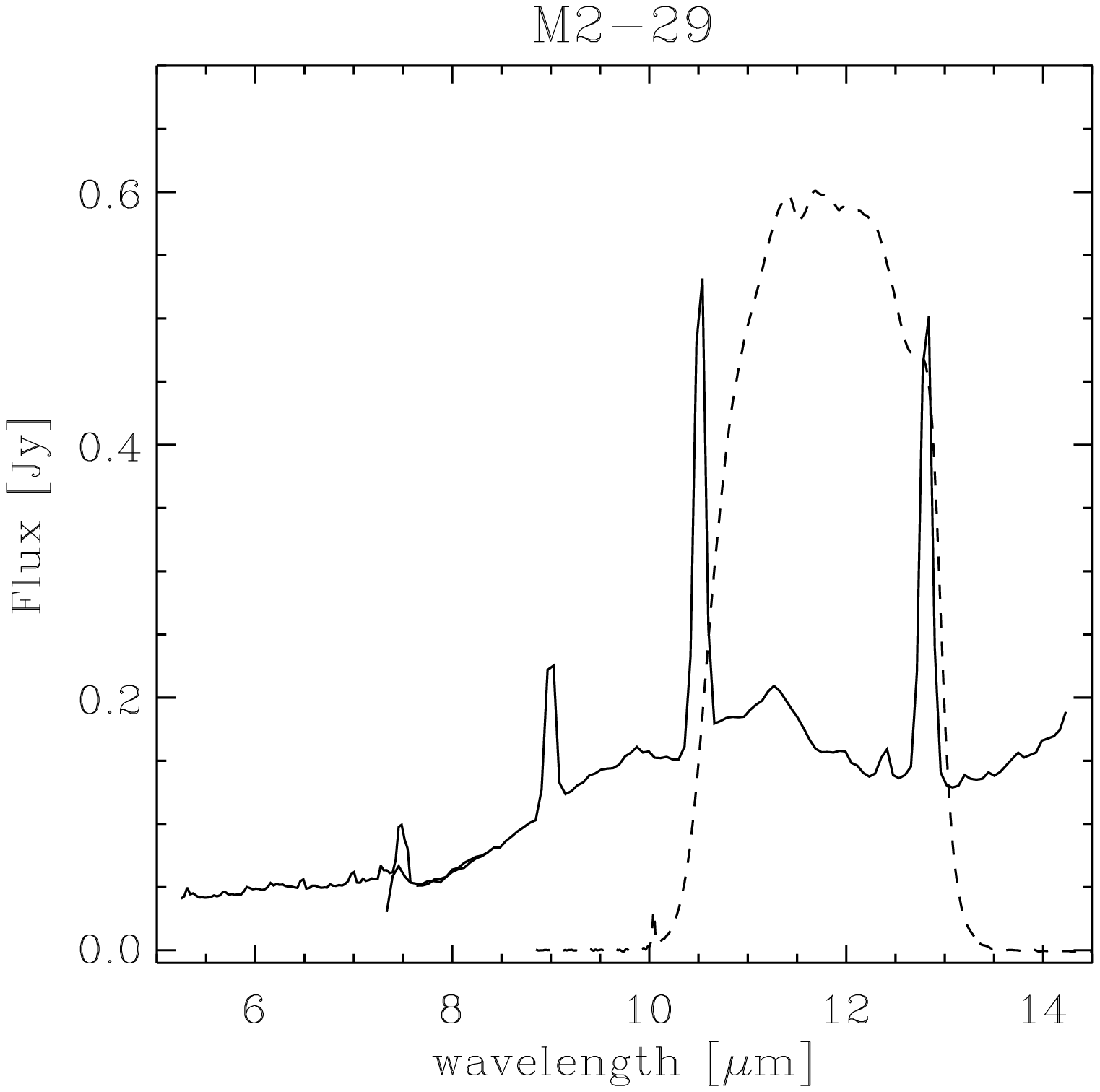}
 \includegraphics{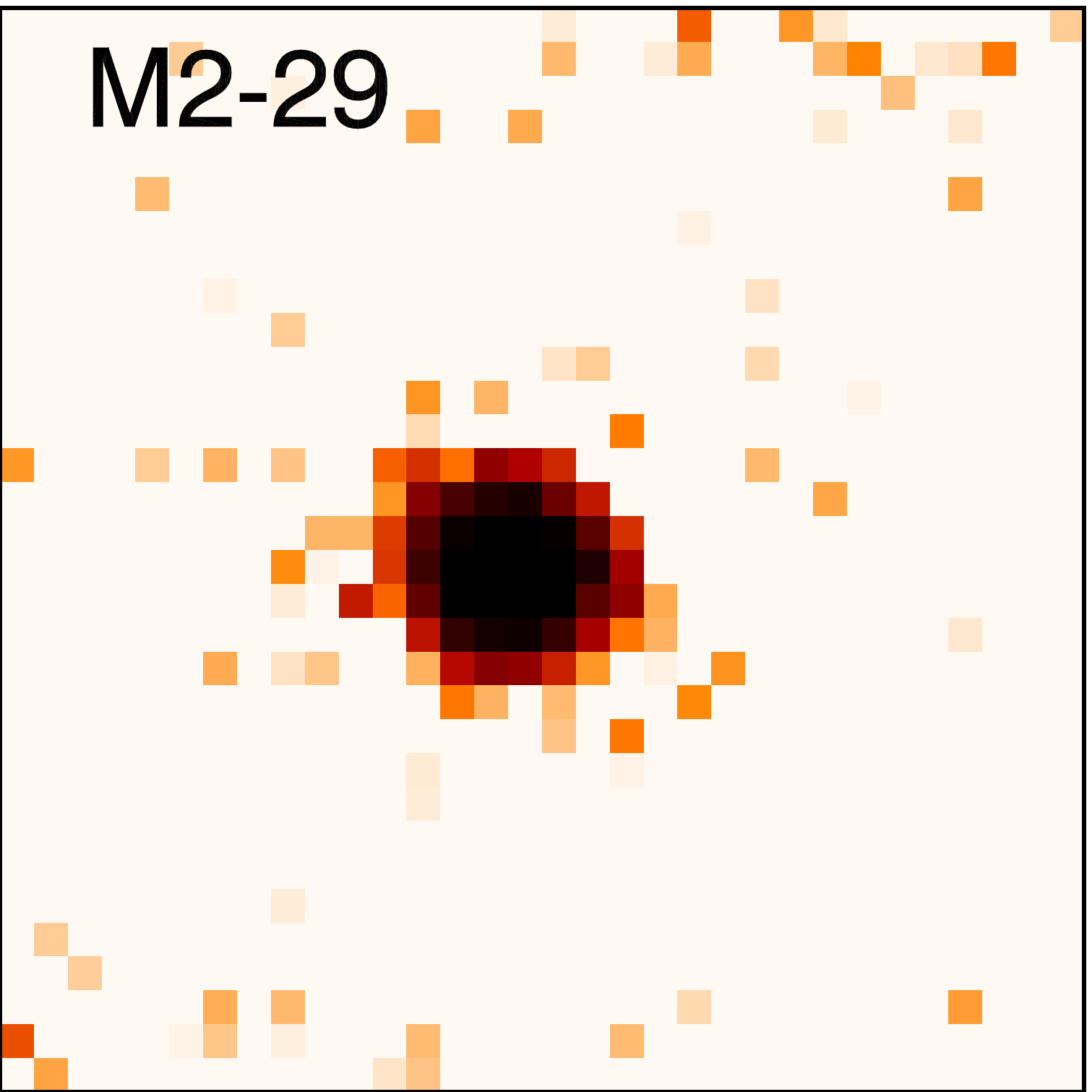}}
      \caption{Left: Spitzer low-resolution spectrum, showing a broad silicate
        band with superposed emission lines and a weak 11.2\,$\mu$m PAH
        band. Right: the VISIR image showing a strong point source. The image
        is $3.8\times 3.8$ arcsec in size. The transmission curve of the
        filter used is shown on the spectrum.  }
    \label{visir_image}
\end{figure}

M\,2-29 is a known dust emitter, detected at 25\,$\mu$m by IRAS (the 12-$\mu$m
detection was rejected for the faint-source catalogue, perhaps because of
confusion with a nearby star). N-band spectroscopy presented by
\citet{Casassus2001} indicates a possible broad silicate emission band at
8-12\,$\mu$m.  We retrieved and reduced archival Spitzer IRS data, which
confirms the presence of silicate emission, and also shows a weak PAH feature
at 11.2\,$\mu$m (Fig. \ref{visir_image}).

To study the location of the dust, we obtained mid-infrared images of M\,2-29
using VISIR on the VLT (Lagage et al. 2004). We used a 0.075 arcsec per pixel
scale and and the VISIR SiC filter ($\lambda_c=11.85\,\mu$m,
$\Delta\lambda=2.34\,\mu$m). The observations were
carried out in visitor mode with great weather conditions, leading to
diffraction limited images during the whole run. The observations were done
using the standard chopping and nodding technique to reduce the background
emission in the mid-infrared.  The data reduction was performed using
self-developed IDL routines described by \citet{Lagadec2008}. Images were
corrected for bad pixels and then co-added to produce a single
flat-field-corrected image, comprising the average of the chop and nod
differences. 

The resulting image at 11.85\,$\mu$m shows an unresolved source
(Fig. \ref{visir_image}), indicating that the emitting source is smaller than
$\sim$0.3 arcsec. Thus, the emitting hot dust only comes from one compact
component (the core), and not from the extended nebula or wing-like feature.

\section{Morphology from HST images}\label{morpho}

The HST H$\alpha$ image \citep{Hajduk2008} shown in Fig.\,\ref{Halpha_image}
presents the complexity of the nebula, including the main ring, the halo, the
wing-like structure, and the unresolved central object. The main ring is
actually composed of two arcs, which may not quite connect.  The gaps between
them are nearly perpendicular to the wing-like structure. There is a slight
indication for elongation or brightening in the EW direction, beyond the gaps
(best seen in the left panel, with a linear scale). The overall impression is
that of a barrel-shaped nebula, with a beginning bipolar flow, a relatively
common shape among planetary nebulae.

The HST H$\alpha$ and [\ion{O}{iii}] 5007\,\AA\ images are almost identical and
show the same extent.  This means that M\,2-29 should be fully ionized
(density bounded).  For the discussion below we use the H$\alpha$ image.

\subsection{The outer halo and dense wind}

\begin{figure}
\resizebox{\hsize}{!}{\includegraphics{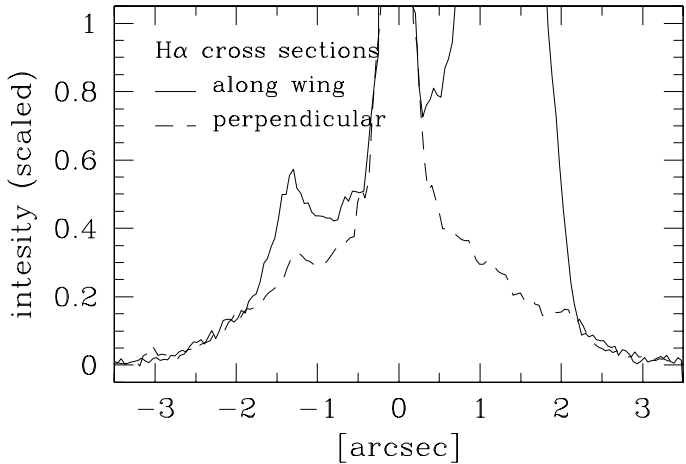} }     
      \caption{The cross-sections through the H$\,\alpha$ HST image of
      M\,2-29 taken along the wing-like structure and perpendicularly to
      it. The intensity scale is the same for both profiles}
    \label{h_jet_holes}
\end{figure}

To inspect  the outer regions, we overplot in Fig.\,\ref{h_jet_holes}
two scaled cross-sections: one extracted along the wing-like structure and the
other perpendicularly to it. The outermost nebular regions decrease
in brightness in the same way in all directions, indicative of  
spherical symmetry. There is no clear evidence
for ISM interaction \citep{Wareing2007}.

Intriguingly, the emission profile extends smoothly and symmetrically inside
the arcs, up to the central object. This is shown in (Fig.\,\ref{h_jet_holes}),
where the small bump on the left hand side corresponds to the arcs. The arcs
and jet appear to be superposed on a largely spherical component whose
brightness smoothly decreases outward.  This component we will call the
``dense wind''. The spherical symmetry of this component points to a
stellar origin. Its brightness at H$\alpha$ image proves it is not
hydrogen deficient.

The emission profile of the dense wind, including the halo, corresponds
roughly to the surface brightness of a sphere filled with matter of a $r^{-0.5}$
density distribution.  This is unusual. The most widely assumed AGB wind
density is $r^{-2}$, as discussed by \citet{VGM2002}. The $r^{-0.5}$ density
implies a mass-loss rate decreasing as $\dot M \propto t^{-1.5}$. 
The surface brightness with an expansion velocity of
10\,km\,s$^{-1}$ gives a current mass loss rate of
$10^{-8}$\,$M_{\sun}$\,\rm yr$^{-1}$.

\subsection{The central nebula}

From the HST images, \citep{Hajduk2008} showed that the unresolved source at
the centre contains an ionized nebula which outshines the central star in the
emission lines.

A number of artificial PSFs were created using the TinyTim v.6.3
\citep{Krist04}, and fitted to this central nebula. Series of sub-pixel images
for different shifts relative to the nebular image were obtained.  The PSFs
were convolved with Gaussian kernels of different FWHMs, and re-sampled to the
resolution of the Planetary Camera. After re-sampling, the charge diffusion
kernel was applied. The PSF was then subtracted from the M\,2-29
image. Minimum residuals were obtained for PSFs convolved with a Gaussian of
$\rm FWHM \sim$ 0.8 pixel. A similar result was obtained for a background
star. The broadening of the PSF with respect to the theoretical one could be
due to an imperfect focus.  From the comparison with the field star, we
constrain the deconvolved FWHM of the nebula as $<0.6$ pixel. This
corresponds, at the distance of 8\,kpc, to a diameter of $< 250$ AU
($3\times 10^{15}$\,cm).

\subsection{The wing-like feature}

To obtain the radial emission profile of the wing-like feature from the HST
image, the other nebular components should be subtracted.

\begin{figure} 
\resizebox{\hsize}{!}{\includegraphics{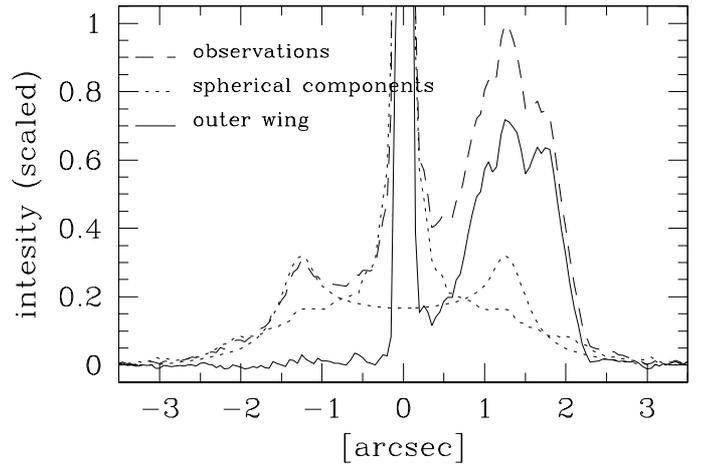} }
\caption{The cross-sections through the H$ \alpha$ HST image of M\,2-29
taken along the wing-like structure. The dashed line shows the observed
intensity profile, the dotted lines show both spherical components - the
arcs (photoionization model) and the dense wind, the continuous line
presents the difference between the observed cross section and the two
spherical components.} 
\label{interp} 
\end{figure}

The profile of the dense wind was obtained from the dashed line in
Fig.\,\ref{h_jet_holes}, by taking the average of this cross section with its
mirror reflection.  The arcs are more difficult, as they are not circular
symmetric and cannot be taken from the perpendicular profiles. Instead we used
a photo-ionization model (see Section 5), approximated by a spherical shell
best fitted to the two arcs, and subtracted the profile calculated from this
model. The unresolved core was not subtracted. At the decomposition weighting
factors were applied.  The result is shown in Fig.\,\ref{interp}.

To the left (south) of the central object, a flat profile was obtained. The
wing-like structure shows up only to the north (right) of the core.

No clear residual is seen at the cross-over point of the wing-like feature and
the arcs.  A superposition on the sky, rather than a physical connection, seems
to be indicated.

\subsection{The arcs}

The successful decomposition of the image cross-section into the dense
wind and the photo-ionization model, shown in Fig.\,\ref{interp},
indicates that the southern arc can be interpreted as a fragment of a
photoionized sphere showing emissivity enhancement at the edge. The same
can be said about the northern arc. A very common interpretation of such
a structure  is that the arcs form the projection of a thin barrel, open
to the two poles. The dense wind fills the inside \emph{and} the outside of the
barrel. This appears to suggest that the density enhancement of the barrel
formed out of the dense wind.

\section{The spectra}

\subsection{Velocity maps}

\begin{figure*} 
\sidecaption
\includegraphics[width=12cm]{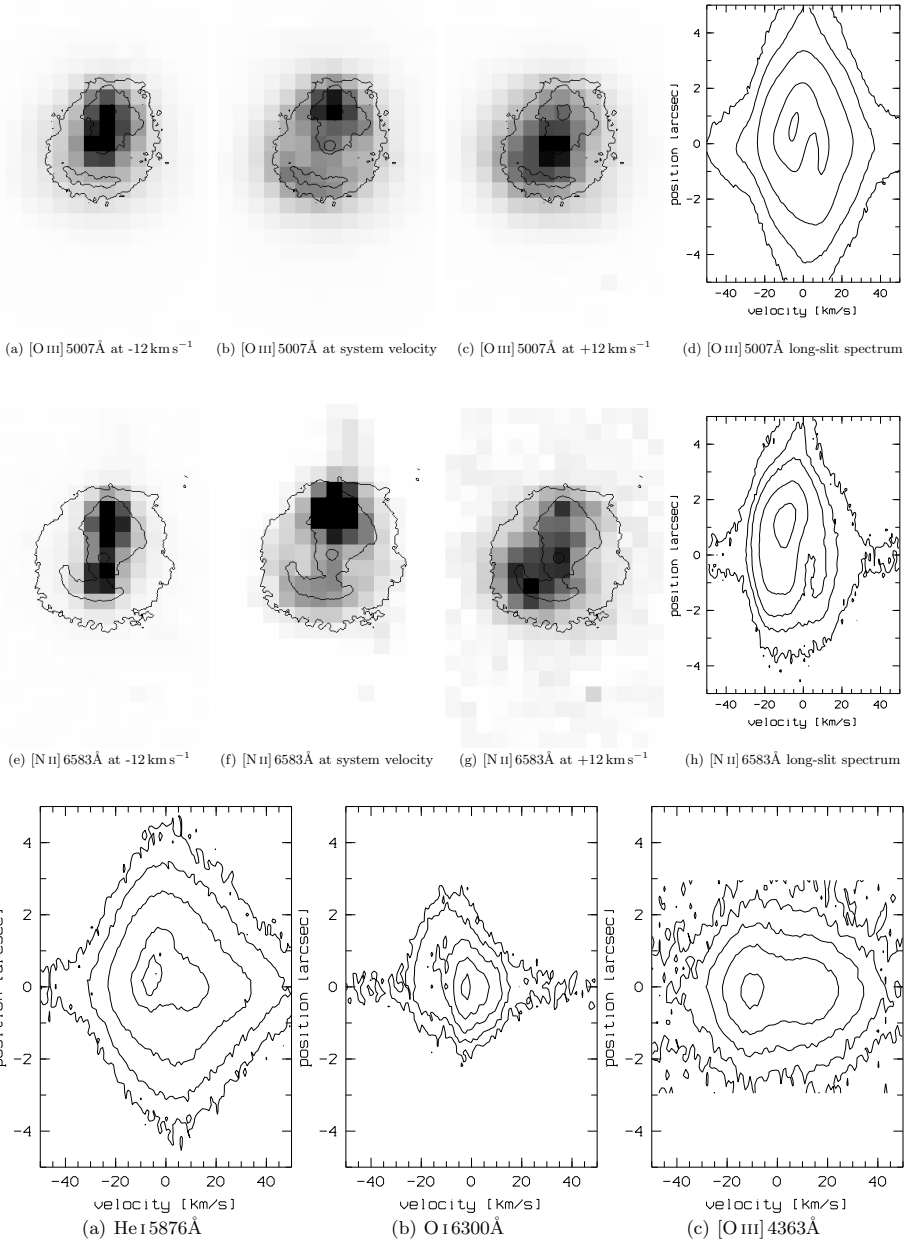}
\caption{The ARGUS-IFU and UVES observations of M\,2-29 in spectral lines of
  [\ion{O}{iii}]\,5007\,\AA\ (upper row) and [\ion{N}{ii}]\,6583\,\AA\ (lower
  row).  The presented slices (left) are extracted from ARGUS data cubes at
  wavelengths corresponding to velocity shifts from the systemic velocity of
  -12\,km\,s$^{-1}$, 0 and +12\,km\,s$^{-1}$ (from left to right
  respectively). The fluxes are shown in grey in linear scale. The black colour corresponds to the relative intensity of 0.9 except of panels (e) and (f) where cuts were applied at levels 0.5 and 0.3 respectively. In both lines the image at -12\,km\,s$^{-1}$ (panels a and e)
  is the brightest. For [\ion{O}{iii}] the central image (b) is of similar
  brightness while (c) is two times weaker.  For [\ion{N}{ii}],
  panel (f) and (g) are fainter by a factor of 3 and 30 with respect to panel
  (e). The  contours are from the HST images, overplotted to the
  same spatial scale; for the [\ion{O}{iii}] slices the [\ion{O}{iii}] image
  is shown while for the [\ion{N}{ii}] slices the H$\alpha$ image is
  shown. The right-most panels show the UVES long-slit spectra where the slit
  was positioned along the wing-like structure i.e. nearly N-S direction. The
  contours are 0.004, 0.02, 0.1, 0.5, 0.95 for [\ion{O}{iii}] in panel (d) and 0.007, 0.03, 0.08, 0.3, 0.8 for [\ion{N}{ii}] in panel (h). }
\label{ifu_6}
\end{figure*}

\begin{figure*} 
\sidecaption
\includegraphics[width=12cm]{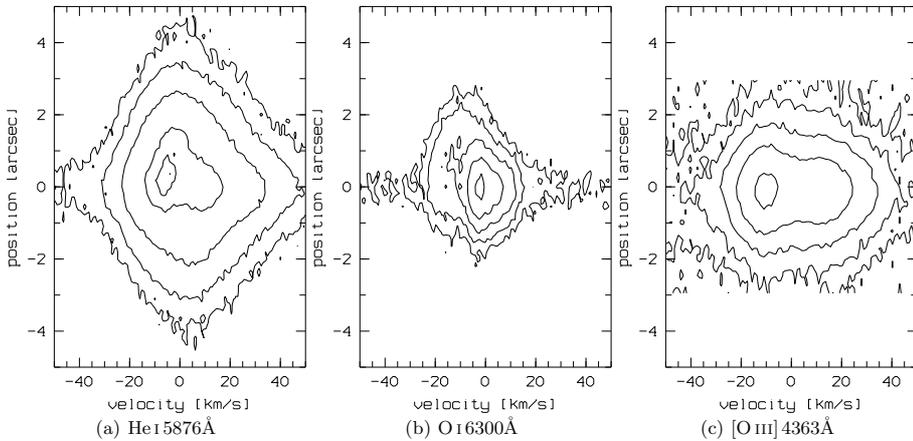}
\caption{The VLT/UVES long slit spectrum of \ion{He}{i} 5876\,\AA\ (left),
  [\ion{O}{i}] 6300\,\AA\ (middle) and [\ion{O}{iii}] 4363\AA\ (right). The
  velocity scale is centred on the systemic velocity. The
  wing-like structure extends up from the centre. The 
    contours  are at levels of 0.02, 0.07, 0.2, 0.7, 0.9 in panel (a), 0.06, 0.1, 0.2, 0.5, 0.9 in panel (b) and 0.01, 0.03, 0.1, 0.3, 0.8 in panel (c). }
\label{vlt_3}
\end{figure*}

In Fig.\,\ref{ifu_6} we present spectroscopic observations in emission lines of
[\ion{O}{iii}]\,5007\,\AA\ and [\ion{N}{ii}]\,6583\,\AA. These are strong
forbidden lines, which probe the inner and outer nebular regions
respectively. For both lines we extracted from the data cubes three
representative slices corresponding to the systemic radial velocity (panels b
and f)\footnote{The velocity is corrected for the radial velocity of the
  object of $-112.2$\,km\,s$^{-1}$ \citep{Durand1998} and barycentre velocity
  for the moment of observations of $-10.4$\,km\,s$^{-1}$ for UVES and
  $-22$\,km\,s$^{-1}$ for ARGUS.} and those shifted by 12\,km\,s$^{-1}$ below
(panels a and e) and above (panels c and g) this value.  The grey scale
represents the relative intensities normalized for each slice separately. The
square pixels are of spatial size of $0.52\times0.52$\,arcsec$^2$. The
contours show the HST images: [\ion{O}{iii}] for the upper row, H$\alpha$ for
the bottom row. The right-most frames show the long-slit UVES spectra, where
the slit was positioned along the wing-like structure i.e. 5\degr\ away from
the N-S direction.

At the central star position, the [\ion{O}{iii}] velocity maps show strong
emission at both the positive and negative velocities (panels a and c) but a
minimum at the systemic velocity (panel b). This emission corresponds to the
unresolved central component and can be interpreted in terms of a bipolar
outflow. The wing-like structure can be traced in panels (a) and (b) with
velocity decreasing outwards; this can also be seen in the long-slit spectra
(panel d). The southern arc is best distinguished at the zero velocity panel
(b).

More details can be derived from [\ion{N}{ii}] images. The wing-like structure
at -12\,km\,s$^{-1}$ (panel e) is clearly more extended than in [\ion{O}{iii}]
at the same velocity (panel a), and is seen at both sides of the central
component. The systemic velocity image (panel f) shows the wing-like structure
extending two arcseconds in the northern direction, far beyond what is seen in
the H$\alpha$ image. The long-slit UVES spectra (panel h) also show this
significant extension.  The southern arc can also been at the systemic
velocity, and  a very weak component is seen in panel (f) extending for a
couple of pixels in S-W direction. Panel (g) shows that a component at
the eastern end of the southern arc is moving away from us, in fact at
a little more than +12\,km\,s$^{-1}$. 

In Fig.\ref{vlt_3} we present more examples of the 2-D (position--velocity)
contours presentation of bright emission lines in the UVES spectrum. All plots
are shown to the same scale, but in the blue spectral region the slit was
shorter so the spatial extent of these lines is smaller.   Many more
emission lines are identified in the spectrum, but the three presented here 
best supplement those in Fig.\,\ref{ifu_6}, showing the variety of line
profiles. 

The emission of the wing-like feature is visible from the centre until about
2\,arcsec up. It is very pronounced in the [\ion{O}{iii}] 5007\,\AA, [\ion{N}{ii}]
(Fig.\,\ref{ifu_6}) and \ion{He}{i} lines, very weak in the [\ion{O}{i}] line and
completely absent in the highly excited [\ion{O}{iii}] 4363\,\AA\ line.

The velocities are relatively low (around 10 km\,s$^{-1}$) compared to typical
expansion velocities of planetary nebulae.

\subsection{Photoionization modelling of the outer nebula}

\begin{figure} 
   \includegraphics{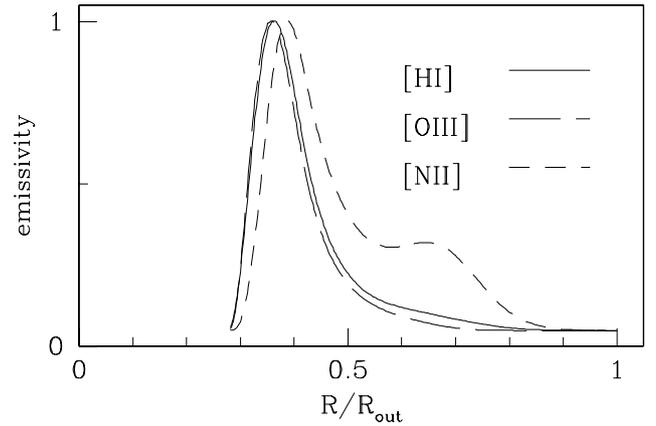}  
   \caption{Model intensity distribution for \ion{H}{i}, \ion{O}{iii} and
 \ion{N}{ii}. The overall density follows H to a good accuracy.}
    \label{model_all}
\end{figure}

To model the spectra and line profiles, we applied the Torun photo-ionization
codes \citep{GZ2003, GZAGGW2006}. The star is assumed to be a black body with
a luminosity and effective temperature. The nebula is approximated as a
spherical shell defined by the radial density distribution and the radial
velocity field. The chemical composition ([O/H]\,=\,7.44) and the observed line
intensities were adopted from \citet{Exter2004}. We have also measured the
line intensities from the low resolution SAAO spectrum \citep{Hajduk2008}.
Table\,\ref{linrat} presents the observed data together with our best model.
The model emissivities are shown in Fig.\ref{model_all}.

\begin{table}
\centering
\caption[]{ Dereddened M\,2-29 emission line flux ratios, relative to
I(H$\beta$)=100.0. Observed data come from our SAAO spectrum  
(dereddened with $c_{\beta}=1.18$ \citep{Hajduk2008}) and from literature   
(dereddened by the authors).} 
\label{linrat}
\begin{tabular}{llrrrr}
\hline\hline
\noalign{\smallskip}
Wavel. & Ident. & \multicolumn{3}{c}{Observed Flux} & Model  \\
    $\left[ \AA \right]$     &        & SAAO & (1) & (2) & \\
\noalign{\smallskip}
\hline
\noalign{\smallskip}
     3726 & [\ion{O}{ii}]   &  43.6 & 42.1  & 34 &  15.8 \\
     3869 & [\ion{Ne}{iii}] &  72.2 & 49.8  & 58 &  19.7 \\
     4363 & [\ion{O}{iii}]  &  27.0 & 14.0  & -  &  1.9  \\
     4959 & [\ion{O}{iii}]  & 150.2 & 146.0 & -  &  46.8 \\
     5007 & [\ion{O}{iii}]  & 449.4 & 424.0 & 452&  134.9\\
     6300 & [\ion{O}{i}]    & 1.6   & 1.1   & -  &  0.4  \\
     6548 & [\ion{N}{ii}]   & 7.0   & 6.9   & -  &  10.2 \\
     6584 & [\ion{N}{ii}]   & 17.8  & 20.8  & 17 &  29.9 \\
     6716 & [\ion{S}{ii}]   & 1.2   & 1.9   & -  &  0.7  \\
     6731 & [\ion{S}{ii}]   & 1.8   &  2.4  & 2  &  0.8  \\
     7136 & [\ion{Ar}{iii}] & 9.0   & 12.5  & -  &  8.3  \\
\noalign{\smallskip}
\hline
\noalign{\smallskip}
\multicolumn{6}{l}{References. (1) \citet{Exter2004}; (2) \citet{Howard1997}.}
\end{tabular} 
\end{table}

\begin{table}
\centering
\caption[]{M\,2-29 photoionization and kinematic model parameters.} 
\label{mod_phot}
\begin{tabular}{ll}
\hline\hline
\noalign{\smallskip}
Parameter & Value \\
\noalign{\smallskip}
\hline
\noalign{\smallskip}
distance &  8 kpc \\
$T_{\rm eff}$ &  70\,000\,K \\
Luminosity  &  1200 $L_{\sun}$  \\
Ionized Mass  &  0.38 $M_{\sun}$ \\
$\log$ He / H + 12 & 10.99 \\
$\log$ N / H + 12 & 7.14 \\
$\log$ O / H + 12 & 7.44$^*$ \\
$\log$ Ne / H + 12 & 6.84 \\
$\log$ S / H + 12 & 5.64 \\
Nebula Expansion Velocity & 12\,km\,s$^{-1}$ \\
Turbulent Velocity Component & 7\,km\,s$^{-1}$ \\
Kinematic Age & 5\,000\,yr \\
\noalign{\smallskip}
\hline
\noalign{\smallskip}
\multicolumn{2}{l}{$^*$ this adopted value might be too low (see the text)}
\end{tabular} 
\end{table}

The one-dimensional model by necessity assumes spherical symmetry.  It solves
for the observed line intensities and the radial surface brightness
distribution. For the latter, we extracted the image cross-sections taken at
45 degrees left and right to the wing-like structure i.e. avoiding the wing
while including the arcs. The two cross-sections are quite similar.  The model
assumes a central cavity: neither the central unresolved nebula, nor the inner
part of the dense wind (i.e. inside the arcs) is included.

The satisfactory solution is summarized in
  Table\,\ref{mod_phot}.  Although the comparison of the H$\alpha$ and
[\ion{O}{iii}] HST images indicates a density bounded PN, the line intensities
(integrated over the whole nebula) are in fact significantly better fitted
when the PN is ionization bounded. We solved this problem by adopting the
density distribution extending far beyond the arcs, obtaining the ionization
front at about 4\,arcsec radius. This makes the model slightly different from
that in \citet{Hajduk2008}. The strong [\ion{O}{iii}] 5007\,\AA\ line is not
well fitted. We could only fit this line by increasing the oxygen abundance by
a factor of 2, still metal-poor  however we did not attempt
to improve the abundance determination.

The model shows a constant expansion velocity with distance from the star, as
derived from the similar widths of lines formed at different radii, broadened
by turbulence.  This is at variance with the usual positive velocity gradient
seen in planetary nebulae. The model turbulence probably reflects the
superposition of different nebular components.  The mass-averaged expansion
velocity derived for M\,2-29 is $12$\,km\,s$^{-1}$ (compared to
$14$\,km\,s$^{-1}$ obtained from [\ion{O}{iii}] only, \citet{GZ2000}). Assuming a
distance of 8\,kpc this velocity results in a kinematic age of about
5000\,yr.

The ionized mass of the nebula is rather high, in view of the conclusion of
\citet{TPDPP1997} concerning old age (and therefore low initial mass) of the
object. The age may have been overestimated, or mass transfer can be proposed.
However, caution should be applied when deriving a mass from a spherical
model for a non-spherical nebula.

\subsection{Extended wings of H$\alpha$}

\begin{figure}
\resizebox{\hsize}{!}{\includegraphics[angle=270, clip=]{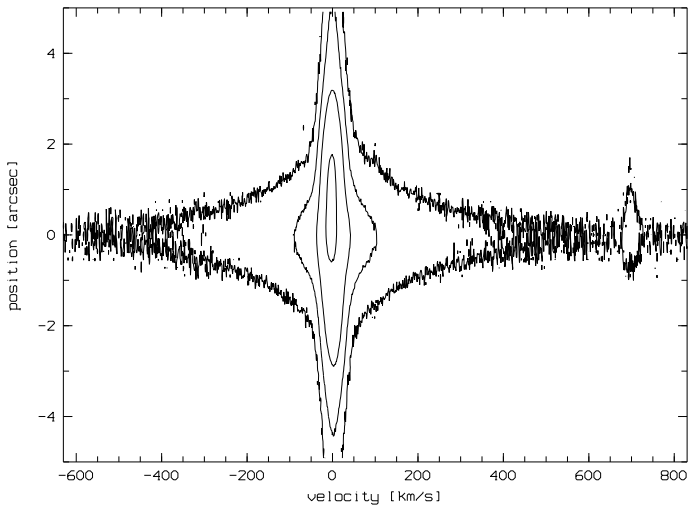} }     
      \caption{The P-V plot of the H$\alpha$ line observed with UVES, with the
        long-slit aligned with the wing-like structure. At the right edge the
        permitted line of \ion{C}{ii}\,6578\,\AA\ can be seen. Contours are
        equally spaced in logarithmic intensity.}
    \label{halfa1}
\end{figure}

The H$\alpha$ emission line reveals very wide wings, seen at the location of
the central unresolved source only. In the UVES position-velocity diagram
(Fig.\,\ref{halfa1}), the vertical structure shows the emission from the
wing-like structure, and the horizontal extent shows the high velocity wings,
located at the same position as the stellar continuum.

\begin{figure}
\resizebox{\hsize}{!}{\includegraphics{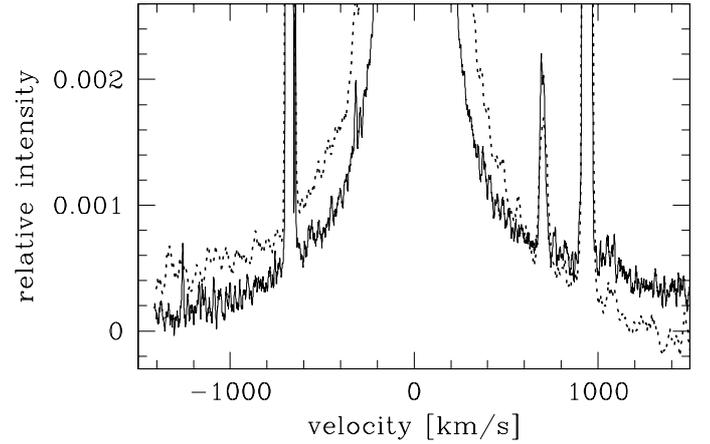} }     
      \caption{The H$\alpha$ line extracted from the central pixels of UVES
        (solid line) and ARGUS-IFU (dotted line). The emissions are normalized
        to the line peak. Note that ARGUS has a lower spectral resolution.}
    \label{halfa2}
\end{figure}

Averaging the slit spectra over the central 1.5 arcsec, the emission wings can
be traced over 1000\,km\,s$^{-1}$ (Fig.\,\ref{halfa2}). The ARGUS-IFU data
cube confirms that at wavelengths beyond $\pm$4\,\AA\ ( $\pm$200\,km\,s$^{-1}$)
from the line centre the emission originates solely from the unresolved
source.

Extended wings can also be seen in [\ion{O}{iii}]\,5007\,\AA\ shown in
Fig.\,\ref{ifu_6}d), and in H$\beta$ (not shown here). For weaker lines it is
not clear whether the wings are present or absent (see e.g. \ion{N}{ii}
6583\,\AA\ line in Fig.\,\ref{ifu_6}h). However, the
\ion{O}{iii}\,5007\,\AA\ line shows wings one order of magnitude narrower than
H$\alpha$. This should be considered when interpreting this emission
component.

The fact that the large broadening is only seen in some lines argues against
electron scattering. The presence of a high velocity component in the central
object is indicated.

\subsection{The permitted emission lines}

\begin{figure}
\includegraphics[width=4cm]{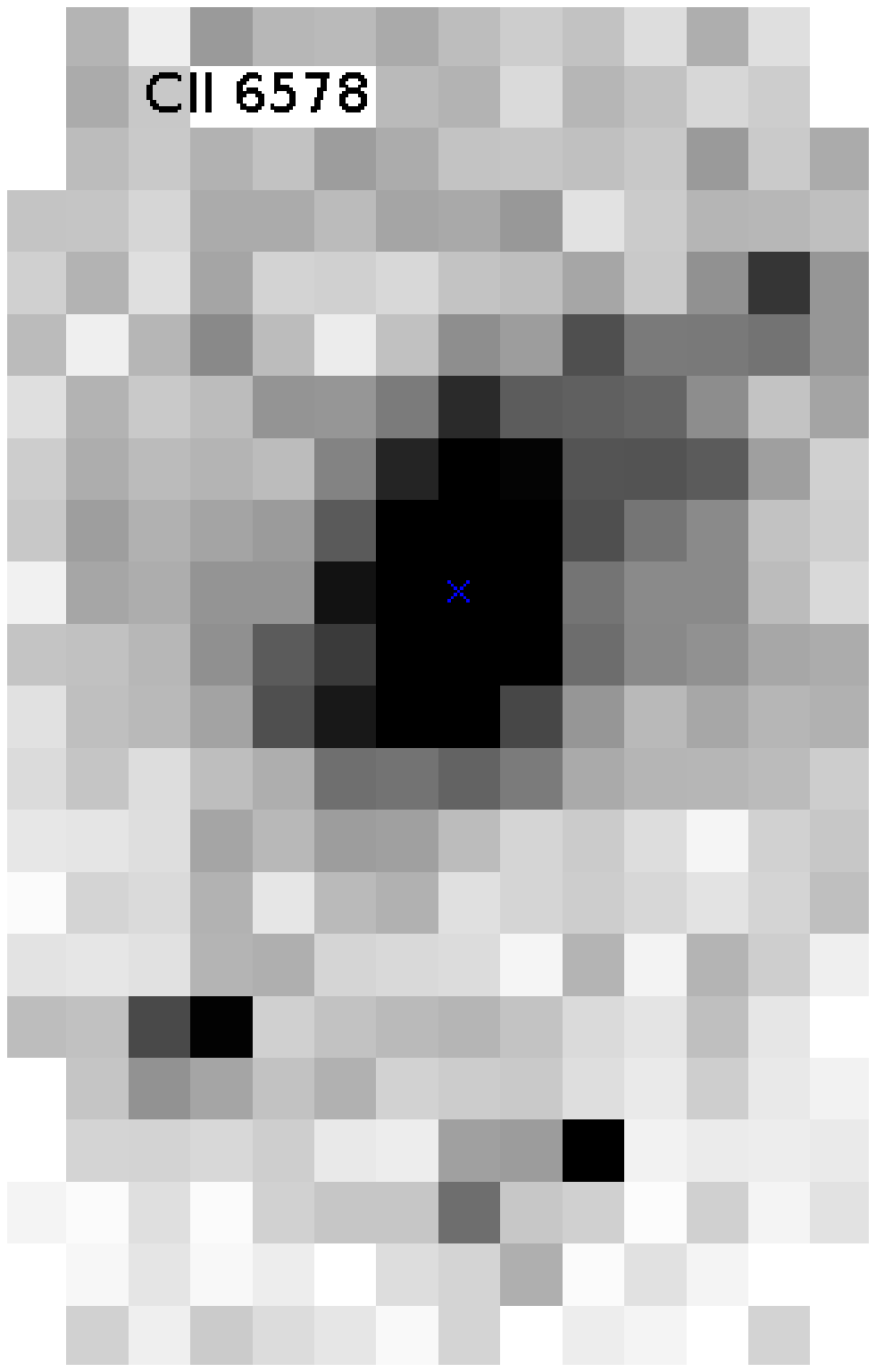}
\includegraphics[width=4cm]{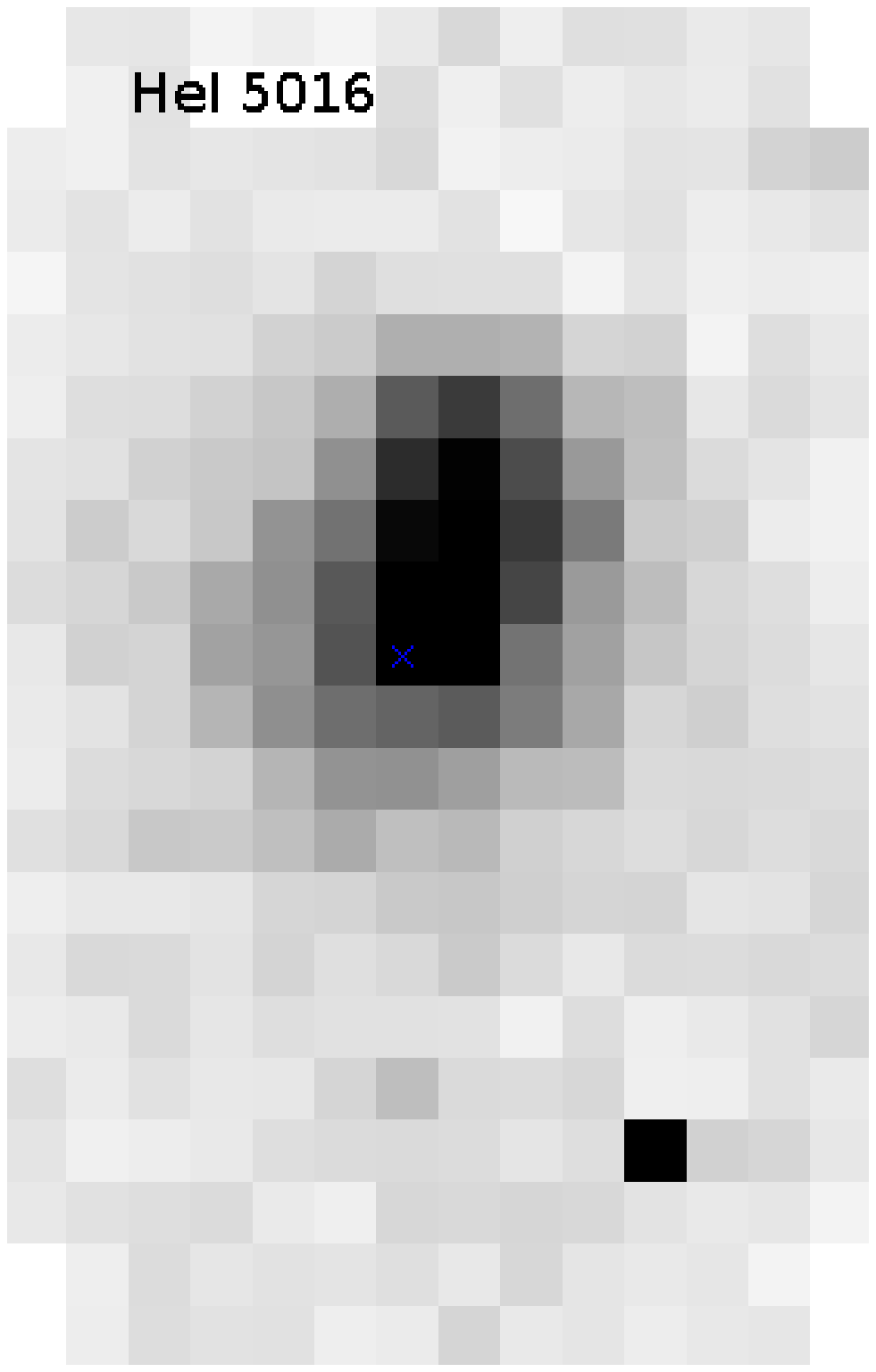}   
      \caption{The ARGUS-IFU collapsed images of the
        \ion{C}{ii}\,6578\,\AA\ and
        \ion{He}{i}\,5016\,\AA\ lines.  North is on top, east
          to the left, the size of a single pixel is 0.52 arcsec.}
    \label{c2_6578}
\end{figure}

The UVES spectrum shows a number of permitted emission lines. The
recombination lines of \ion{C}{ii} at 4267.26 and 6578.05\,\AA\ are present
(e.g. Fig.\,\ref{halfa1}).  Also the recombination lines of \ion{O}{iii} at
4638.85, 4641.81, 4649.14 and 4650.84\,\AA\ can be identified. These lines are
useful in abundance analysis, e.g. \citet{garcia2005}.

For the well exposed \ion{C}{ii}\,6578\,\AA\, line, an ARGUS-IFU image was
obtained by integrating over its line width.  This line originates in an
extended region as can be seen in Fig.\,\ref{c2_6578} where it is compared
with \ion{He}{i}\,5016\,\AA\ line. Concerning the other permitted lines the
situation is not so clear since they are weaker and more noisy. The well-known
group around 4650\,\AA\ falls at an echelle order merging where distortions
may occur; it is also not covered by the ARGUS-IFU spectral range. The UVES
data suggests that it may also form in an extended region but this requires
confirmation.

We also identified a number of permitted \ion{N}{iii} lines (4097.33, 4103.43,
4634.14, 4640.64, 4641.85\,\AA). They are often interpreted in terms of the
Bowen mechanism (see e.g. \citet{selvelli2007}), where \ion{He}{ii} Ly$\alpha$
overlaps with an \ion{O}{iii} line, and the resulting \ion{O}{iii} cascade itself has a
frequency coincide with an \ion{N}{iii} resonance line. The \ion{O}{iii} Bowen lines
below 4000\,\AA\ fall within a very noisy area of our VLT spectrum and can not
be identified.  The very weak \ion{He}{ii} line argues against a Bowen
mechanism in the nebula of M\,2-29.  \citet{Mihalas1971} proposed dielectronic
recombination for explaining the \ion{N}{iii} line complex near
4650\,\AA.  The \ion{N}{iii} lines are only detected from the central
nebula.

\subsection{Interstellar absorption and distance}

\begin{figure}
\resizebox{\hsize}{!}{\includegraphics{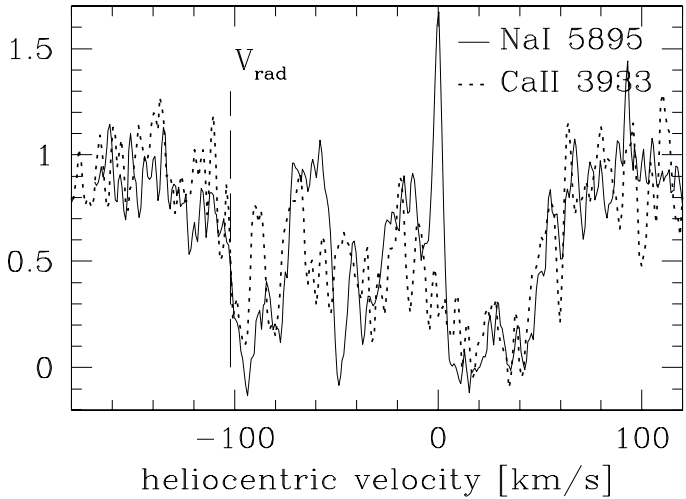} }     
      \caption{The interstellar absorption line profiles of ions \ion{Na}{i} 
and \ion{Ca}{ii}\,K in the spectrum of M\,2-29. The profiles are shown in
 the heliocentric reference frame.}
    \label{NaCa}
\end{figure}

The UVES spectra of M\,2-29 show interstellar absorption lines of \ion{Na}{i} 
and \ion{Ca}{ii}, shown in Fig.\,\ref{NaCa}. 

The equivalent width (EW) of \ion{Ca}{ii}\,H and K lines 
  (i.e. adding together all Doppler components) shows an approximate
linear relation with distance \citep{megier2005}, as derived from OB stars
with Hipparcos parallaxes. For M\,2-29, the measured EW of the full
\ion{Ca}{ii}\,K absorbing complex is about 1500\,m\AA. (The H-band is confused
with H$\epsilon$.)  This results in a distance of 4 or 6 kpc when applying
the relations of \citet{megier2005} or \citet{gazinur2005} respectively. It is
not clear whether the linear relation can be extrapolated into the inner
regions of the Galaxy; however, the indicative values are in agreement with a
Galactic Bulge location.

The interstellar absorption line profiles show components from different
interstellar clouds. The three main components are at (heliocentric) velocities of
0--50\,km\,s$^{-1}$, at $-$60 to $-20$\,km\,s$^{-1}$, and at $-$100 to
$-$80\,km\,s$^{-1}$. The absorption at positive velocities can be associated
with the Molecular Ring around the inner Galaxy, while the intermediate
negative velocity component is consistent with the 3\,kpc expanding ring as
seen in CO \citep{Dame2001}. However, the Galactic CO maps show no counterpart
to the most negative velocity component in M\,2-29. This component cannot be
explained with Galactic rotation, nor Bulge rotation\footnote{The mean
  rotation of the Galactic bulge at the longitude of M\,2-29 is around
  +50\,km\,s$^{-1}$ \citep{Minniti2008}.}.

HIPASS spectra in this region do show an \ion{H}{i} emission feature
corresponding to this negative velocity component. It may be related to a
nearby catalogued high velocity cloud, HVC 002.9-06.2-088 \citep{Putman2002},
with the same velocity, as the emission at this velocity seems to extend over
some distance. The distance to this cloud or complex is not known, but as
M\,2-29 is clearly located behind it, we suggest it is situated relatively
close to the Galactic centre. Small \ion{Ca}{ii} and \ion{Na}{i} absorption
clouds with similar high velocities appear to be common in the Galactic halo
\citep{Bekhti2008}, with a variety of origins.

It is remarkable that the non-rotational interstellar absorption component
near $-$100\,km\,s$^{-1}$ is very close to the systemic radial velocity of
M\,2-29.  As a PN, or related object, M\,2-29 is expected to be old and evolved,
and any association with the interstellar component is deemed accidental.

\section{Disentangling the nebula}

The nebula of M\,2-29 shows four distinct components: the inner unresolved
nebula, the dense wind extending outwards, the arcs, and the wing-like
structure. The arcs appear to be an enhancement within the dense wind rather
than an independent structure. Here we  attempt to obtain an understanding of 
the other two structures. 

\subsection{An outer partial ring: the wing-like structure}

The 'wing-like' feature is a low-velocity structure.  Based on the observed
location and the lack of interaction with the arc, we locate it outside of the
main nebula.  We identify it as a partial, expanding ring, surrounding the
main nebula, and  whose plane is almost coincident with the
  line of sight. The ring has an inner radius of about 2 arcsec.

The velocity structure along this ring is depicted in Figs. \ref{ifu_6} and
\ref{vlt_3}. In the [\ion{O}{iii}] line, only the top (north) of the ring is
seen, but the [\ion{N}{ii}] line shows the more complete structure.  In the
southern part, the [\ion{N}{ii}] is displaced to the left (see
Fig. \ref{ifu_6}g).  The displacement suggests that the plane
  of the ring is seen at an inclination of approximately 20 degrees with
  respect to the line of sight. Because the displacement is seen in the rear
  part of the ring it is not unlikely that the structure is warped.

The line splitting shows that it is expanding: an expansion velocity (with a
minor correction for the inclination) of about $12\pm2$\,km\,s$^{-1}$ is
indicated.

The UVES slit was aligned with the brightest part of this structure, however
because of the inclination the southern extent fell outside the slit. This is
made clear by the ARGUS image. The position-velocity diagrams from the UVES
data therefore have an incomplete coverage. Still, the brightness distribution
of the ring is strongly asymmetric, while the velocities appear symmetric.
The peak [\ion{O}{iii}] emission occurs at a velocity of $-10$\,km\,s$^{-1}$,
offset by about 1 arcsec from the star. This corresponds to a location on the
ring about 30 degrees from the line of sight to the centre. The bright
[\ion{O}{iii}] emission occurs over an arc of roughly 90 degrees, on the
forward side of the ring. Some faint emission originates from the backside,
seen towards the southwest at positive velocities, in Fig. \ref{ifu_6}c.

The bright [\ion{N}{ii}] emission covers a much wider arc, extending for some
120 degrees. At a much fainter level the ring may be almost closed. The
[\ion{N}{ii}] emission is spatially also much more extended than
[\ion{O}{iii}], with a fainter outer region with a radius of at least 4
arcsec. This is seen in Fig. \ref{ifu_6}f as the northern extension. This
region only contributes at near-systemic velocity, suggesting that the
expansion velocity decreases sharply from 2 to 4 arcsec radius. The radial
extension is not seen towards the south, showing that at its outermost radius the
ring is is not closed, perhaps limited to some 180 degrees.

The ring is also seen in \ion{He}{i}, \ion{C}{ii}, and [\ion{O}{i}], at the
location of the brightest [\ion{O}{iii}] and [\ion{N}{ii}] emission.  Thus,
there is sufficient column density in this direction to cause ionization
stratification and an ionization boundary, as also indicated by the
photo-ionization model. The lack of [\ion{O}{iii}] beyond 2 arcsec is
consistent with the photoionization model of Fig. \ref{model_all} .

Finally, this ring system is located close to the equatorial plane
 of the main nebular barrel defined by the northern and
  southern arc (see Sect.\,\ref{morpho}). A slight misalignment is explained
by the inclination.

Thus, we arrive at a model of an expanding equatorial, partial ring, 2--4
arcsec in radius, with a strong azimuthal density gradient, a radial density
gradient, and an expansion velocity of $12\pm2$\,km\,s$^{-1}$ at 2 arcsec,
decreasing with increasing radius.

\subsection{The unresolved disk}

The central source is detected in emission lines of H$\alpha$, [\ion{O}{iii}]
4363 and 5007\,\AA, \ion{He}{ii}, the Bowen \ion{N}{iii} lines, and
[\ion{O}{i}].  Other lines are present but are confused by the spatial and
velocity overlap with the extended emission.  The very high velocity wings,
and the [\ion{O}{iii}] 4363\,\AA\ line are \emph{only} seen in this core.  The
size of the core is $<0.03$ arcsec, or $<250$AU at the distance of the
Galactic Bulge.

The [\ion{O}{iii}] 4363/5007 line ratio is much higher in the core than in the
extended nebula.\footnote{In the photoionization model of the outer nebula the
  [\ion{O}{iii}] 4363\,\AA\ line is underestimated by an order of magnitude
  which is another argument that it does not belong to the outer main ring.}
This indicates high density and temperature, in agreement with
\citet{TPDPP1997} who first showed the presence of a high density and high
electron temperature component close to the star. We ran some test models to
estimate the parameters of the inner nebula, using the central star parameters
derived above. In these models, the ratio 4363/5007 [\ion{O}{iii}] lines can
be reproduced with $T_{\rm e} \approx 1.4\times 10^4$\,K and $n_{\rm e}
\approx 6\times 10^5\,{\rm cm}^{-3}$.  These values are a little lower than
the estimates of \citet{TPDPP1997}.  At these densities, the mass of the
compact nebula needed to reproduce the observed H$\beta$ flux is
$10^{-4}$\,$M_{\sun}$. This value is not very accurate, as the core is
unlikely to show constant density and spherical symmetry, but provides a
believable order of magnitude.  Other lines often used as the
  electron density indicators, e.g. lower ionized/excited [\ion{O}{ii}] or
  [\ion{S}{ii}], predominantly originate in the wing-like structure. This
  makes an extraction of the central component much more unreliable, in no
  case as clear as the [\ion{O}{iii}] 4363\,\AA\ line.

All lines show velocity broadening of $\sim 10$\,km\,s$^{-1}$.  For this
velocity, the kinematical age of the core is  less than 100
years, requiring a continuing mass-loss rate of $\sim 10^{-6}\,M_{\sun}\,\rm
yr^{-1}$. The lower limit to the mass-loss rate is 1 to 2 orders of magnitude
lower than that required for the extended nebula (assuming a kinematic age of
5000\,yr), and the surface brightness distribution of the extended wind indeed
implies a linearly decreasing mass loss rate with time. It raises the
question, however, why the mass-losing star is not seen. Such mass-loss rates
and velocities are found in red giants. (The wind from the hot central star
will be very much faster). Neither the spectrum nor the photometry shows
evidence for a contribution from any other star than the PN central star.  The
2MASS photometry excludes the presence of a red giant more luminous than
100\,$L_{\sun}$. Although a mass-losing star has clearly been present in the
recent past, it does not appear to be present at the moment.

The alternative to on-going stellar mass loss is a stable, rotating structure.
This has some evidence in favour: the line widths decrease with decreasing
excitation, from $\pm 18$\,km\,s$^{-1}$ HWHM for [\ion{O}{iii}] line to close
to a few km\,s$^{-1}$ or less for the [\ion{O}{i}] line consistent with
rotation velocities decreasing outward. \citet{Hajduk2008} derive a
 putative orbital period for the  wide
binary of $\sim 17$\,yr, implying an orbital velocity of approximately
10-15\,km\,s$^{-1}$. They also show that a circumbinary dust disk in such a
system can explain the long-duration eclipse.  A rotating, circumbinary disk
would show an innermost velocity a little less than the binary orbital
velocity, at some 10\,AU, in agreement with [\ion{O}{iii}]. The [\ion{O}{i}]
velocity corresponds to radii of a few hundred AU.

The extreme H$\alpha$ wings show the presence of a fast outflow, which in the
presence of a stable disk will have a bipolar flow pattern. The dense wind
shows the presence of outflow: we interpret this as a disk wind, caused by
evaporation of the photo-ionized gas. The limit of stability of the gas disk
is set when the thermal velocities exceed the rotational velocity. (Beyond
about 100\,AU, only neutral gas or dust can be stable.) Evaporating gas can be
accelerated by the fast H$\alpha$ wind, as a mass-loaded wind
\citep{Borkowski1995}. The [\ion{O}{iii}] core emission shows a double peak,
and this may indicate it forms in part in such a bipolar wind.

A likely interpretation of the inner nebula is, therefore, in terms of an
opaque disk seen edge-on \citep{Hajduk2008} with ionized bipolar outflows.
The infrared dust emission and neutral [\ion{O}{i}] line originate at the
outer static boundary.  The dense wind originates in the evaporating 
disk.  At larger distances such an evaporating-disk wind is 
expected to be essentially spherically symmetric \citep{Alex2008}. The
disk can be located in the same plane as the outer partial ring, at an
inclination of 20 degrees from the line of sight, with the bipolar flow
directed towards the openings between the two arcs.

\section{Discussion}

\subsection{The  nebula}

The one-side structure of the wing in M\,2-29 is unusual but not unique. It is
duplicated in He\,2-428, and in A\,79 \citep{Rodriguez2001}, and possibly also
in the symbiotic star V417 Cen \citep{vanWinckel1994}. In all four cases, the
structure is essentially identical, allowing for the difference in viewing
angle. But no further object with a similar morphology could be identified in
the IAC morphological catalog \citep{Manchado1996}, confirming the rarity of
this structure.

Outer asymmetries can be formed by interaction with the interstellar medium,
and these are a very common phenomenon \citep{Wareing2007}. In M\,2-29, there
is indeed a possible association with interstellar gas at a similar velocity,
and a possible ISM sweep-up should be explored. However, the good alignment of
the outer ring with the equatorial plane of the nebula, the presence of
expansion, and the existence of a few very similar objects, all suggest that
the partial ring was part of the planetary nebula ejection.

The arcs show the presence of a more (if not entirely) spherically symmetric
component in the nebula, located within the outer ring. This suggests a
scenario where the mass ejection was initially highly asymmetric, but became
less so over time. The arcs could  trace a temporary mass-loss enhancement,
or they could be caused by the pressure enhancement from the onset of
ionization.  The compact core is evidence that a fraction of the ejected mass
was retained close to the star.

The ejection of the partial ring can be related to an interaction with a
companion with an elliptical orbit.  Such an interaction is discussed by
\citet{Soker2001}, their scenario 4. If the orbit is such that interaction
occurs at periastron only, amplifying or initiating the mass loss at this
point, the mass loss will occur preferentially in the direction of the
velocity vector of the mass-losing star at periastron, and not at the systemic
velocity. As the stellar radius increases, the binary orbit may circularize or
the stellar mass loss may become self-sustaining: the mass loss will now occur
everywhere along the orbit, centred on the systemic velocity.

The interaction between these two winds offset in velocity \citep{Soker2001}
provides a natural explanation why an identical structure is seen in several
objects, and seems the most plausible explanation for the outer wing, and the
fact that the asymmetry occurred only during early mass
  loss.  There is some evidence that V417\,Cen,  which shows a
  similar morphology, has an elliptical orbit \citep{vanWinckel1994}.

\subsection{Classifying the star}

M\,2-29 is one of only three PNe known to show dust eclipses 
  (the others are: NGC\,2346 and CPD$-56\degr8032$ \citep{Hajduk2008, Cohen2002,
    deMarco2002}), and one of two with a binary period of order 1 month
 (NGC\,2346 has period of 16 days \citep{Hajduk2008,
    deMarco2009}).

It is also one of few showing a dense, unresolved nebular core inside an
extended PN.  Other cases include EGB\,6, a very faint and nearly circular PN
around a  binary composed of a very hot white dwarf and a
  visual point-like companion with dense emission-line nebula
  \citep{Bond1993}. \citep{Rodriguez2001} finds a similar core in He\,2-428.
B[e] stars, a mixed class which includes some young PNe \citep{Lamers1998},
also show compact cores and H$\alpha$ line wings similar to M\,2-29
\citep{Zickgraf2003}. But M\,2-29 lacks the low-ionization iron lines found in
B[e] stars. A very late thermal pulse leads to the ejection of a new core
\citep{vanHoof2007}, but this core is hydrogen-poor and different from what is
seen in M\,2-29.

A relation to the dusty symbiotic stars has been proposed \citep{Hajduk2008,
  Miszalski2009}, based on the orbital period of M\,2-29.  Symbiotic systems
contain a mass-losing giant, and a hot white dwarf which irradiates the wind:
the ionized gas shows dense cores (e.g. M\,2-9, OH\,231.8+04.2). But although
possibly related, M\,2-29 does not appear to be a symbiotic star. It shows
neither the Raman lines at 6830 \&\ 7082\,\AA, nor any indications for a cool
(giant or yellow) companion in the spectrum. In the case of M\,2-29, although
there is evidence for two companions \citep{Hajduk2008}, neither appears to be
the required mass-losing giant.

However, M\,2-29 will likely be a symbiotic star at some time. If the
companion is a white dwarf, the object would have been a symbiotic star or
symbiotic Mira in the recent past, while losing mass. If the companion is main
sequence, it will become a symbiotic once the companion ascends the giant
branches \citep[e.g.][]{deMarco2009}.

\subsection{Evolution of the disk}

Compact dust disks are common around post-AGB stars \citep{deRuyter2006}, and
are increasingly being identified around the central stars of planetary
nebulae: examples are the ant nebula \citep{Chesneau2007}, the helix nebula
\citep{Su2007}, CPD\,$-56\degr8032$ \citep{Cohen2002}, M\,2-48 \citep{Phillips2008}
and NGC\,2346 \citep{Costero1986}. 

The origin and evolution of these disks is not yet understood.  They are
interpreted  as circumbinary disks which formed during the
  mass loss phase, or as debris disks either left over from the main sequence
  phase or built recently from tidally disrupted smaller orbiting bodies as
  discussed in e.g.  \citet{Su2007} or \citet{Brinkworth2009}.  The plethora
of different classes of objects with such disks (B[e] stars, symbiotic stars,
RV\,Tau stars) shows that they form relatively easily. Binary companions
appear required in all cases, but the orbital periods range from 200 days
(post-AGB binaries with disks) to 100\,yr (such as M\,2-9).

Interestingly, similar dust disks have not been discovered around the
post-common-envelope binaries seen in about  15--20\%\ of PNe
  \citep{Miszalski2009}, with periods of hours to days. 
  Although models \citep{Sandquist1998, NorBla2006} indicate that the
  common envelope is not necessarily completely ejected and there remains some
  gas around the system, there is no evidence that this residual leads to a
  dust disk. The  known dust disks are seen around longer-period systems
  which avoided a common envelope phase.

M\,2-29 is argued to trace the phase where the gas disk is lost and the dust
disk remains. Rotation velocities far exceed thermal velocities while the gas
is molecular or neutral. The dust will settle towards the mid plane at this
time \citep{Dominik2003}. But once ionization starts, the thermal velocities
are comparable to the rotation velocities, and a disk wind initiates. In
M\,2-29, the disk wind began $\sim 10^3$\,yr ago, but has been decreasing over
time, likely reflecting the decreasing gas reservoir. At the current time,
some $10^{-4}\,M_{\sun}$ of gas remains.  The current disk-wind mass-loss rate
of $10^{-8}\,M_{\sun}\,\rm yr^{-1}$ gives a decay time of the disk similar to
the age of the nebula.
 
The future evolution is of interest to the disks seen around more evolved
objects. The Spitzer spectrum with the strong silicate emission and PAH
emission indicates that the dust is relatively hot. The dust will cool sharply
once the star enters the white dwarf cooling track where it rapidly fades by a
factor of 100. The central star of the helix is such a cooling track star.
The cooler dust will emit mainly at longer wavelengths: the dust disk of the
helix is detected mainly at 24\,$\mu$m and 70\,$\mu$m \citep{Su2007}. Such cool
disks can have gone largely unnoticed.

\begin{table}
\caption[]{\label{ages}Ages and dust masses of the disks in planetary nebulae
  }
\begin{flushleft}
\begin{tabular}{lrrl}
\hline\hline
\noalign{\smallskip}
  Nebula & Nebular Age & Dust Mass & Reference \\
       &     [yr]  &   [$M_{\sun}$]  \\ 
\noalign{\smallskip}
\hline
\noalign{\smallskip}
CPD$-56\degr8032$ &  $10^2$         &  $3 \times 10^{-4}$ & 
(2), (3) \\
Ant nebula (Mz 3) & $10^3$         &   $1 \times 10^{-5}$ & 
(1), (4) \\
M\,2-29          &  $5\times10^3$   & $ 10^{-6}$          &  This paper \\
Helix nebula     &  $1.1 \times 10^4$ & $4 \times 10^{-7}$  & 
(5), (6)\\
\noalign{\smallskip}
\hline
\noalign{\smallskip}
\multicolumn{4}{l}{References. (1) \citet{Chesneau2006}; (2) \citet{Chesneau2007};}\\
\multicolumn{4}{l}{(3) \citet{deMarco1997}; (4) \citet{Guerrero2004};}\\
\multicolumn{4}{l}{(5) \citet{Meaburn2008}; (6) \citet{Su2007}.}
\end{tabular}
\end{flushleft}
\end{table}

Table \ref{ages} lists the ages and disk dust masses for four planetary
nebulae where these have been measured.  The ages are defined from the end of
the AGB, i.e. the ejection of the main nebula. The dust masses are taken from
the given references, where we assumed a gas-to-dust ratio of 100 for
M\,2-29. The age of CPD\,$-56\degr8032$ has likely been underestimated as the
object has been known for longer than this \citep{Gill1900}.

The Table indicates that the disk masses decrease sharply as the star ages.
Although this should be taken with care because of the small sample and the
possibly disparate origins of the objects, it appears possible that much of
the dust mass is lost during the PN phase. Dust can be carried away in the
wind from the evaporating gas disk, and unshielded dust may be destroyed by
the intense UV radiation field. \citet{Chesneau2006} suggests that an
efficient dissipating process is occurring in the disk of CPD\,$-56\degr8032$.

 The original masses of the disks are unknown. However, disks around post-AGB
stars show dust masses similar to CPD$-56\degr8032$: RU Cen and AC Her show
$5 \times 10^{-4}$ and $2 \times 10^{-4}\,\rm M_\odot$ respectively \citep{Gielen2007}.
These objects have  stars which are still too cool to cause ionization, and 
have avoided ionization-driven dissipation.

Thus, we suggest that the disk of the helix nebula may be the remnant of a
post-AGB dust disk, rather than a debris disk as preferred by \citet{Su2007}.
Once the star is on the cooling track, the disk becomes largely stable and the
further dissipation time is set by Poynton-Robertson drag, with a time scale
of order $10^7$\,yr. Thus, the disks may remain detectable for the order of
$10^8$\,yr, long after any sign of the planetary nebula has disappeared. 

Several disks have been discovered around white dwarfs
\citep[e.g.][]{Gaensicke2006}. Surveys for such disks around the oldest white
dwarfs have been unsuccessful \citep{Kilic2009}, but around 15\%\ of local
white dwarfs show metal-rich material in their photosphere, indicative of
accretion from a residual dust reservoir \citep{Sion2009}. Although these may
have a variety of origins, they may include the progeny of the M\,2-29-type
disks.

\section{Conclusions}

We have presented a detailed study of one Bulge planetary nebula, with the aim
to find evidence for disk evolution. Our main findings are:

\begin{enumerate}

\item M\,2-29 has a likely distance consistent with membership of the Galactic
  Bulge. It shows absorption features of a cloud with high velocity,
  which we tentatively identify with the nearby HVC 002.9-06.2-088.

\item The nebula is surrounded by a partial ring, slowly expanding at $12\,\rm
  km\,s^{-1}$. Such a structure may be derived from interaction with the
  interstellar medium, but based on the expansion, and the alignment with the
  inner and outer nebula, we associate it with an early mass-loss phase from
  the central star. A model where mass loss is triggered initially only during
  periastron of a companion in an elliptical orbit can explain the structure:
  in this model, the ring was ejected with a preferential velocity in the
  orbital plane, reflecting orbital motion of the central star.

 \item We find evidence for on-going mass loss from the central source,
   decreasing with time, with a slow expansion velocity of $10\,$km\,s$^{-1}$,
   and with a current rate of $\sim 10^{-8}\,M_{\sun}\,\rm yr^{-1}$. There is
   however no evidence for any mass-losing star which could be the source of
   this material.

 \item There is evidence for a very fast wind ($\sim 10^3\,\rm km\,s^{-1}$)
   from the central star, based on the H$\alpha$ line wings. The
   [\ion{O}{iii}] also shows wings, but a factor of ten narrower.

 \item The unresolved core with diameter $<250\,$AU shows strong forbidden and
   permitted emission lines with a range of ionization states, down to
   [\ion{O}{i}].  The velocity HWHM of the 4363[\ion{O}{iii}] line is
   18\,km\,s$^{-1}$; [\ion{O}{i}] is much narrower. Because of the very short
   dynamical time scale, and the velocity gradient with excitation, we
   interpret this as a stable, rotating system. The mass of the core is
   estimated as $M_c \sim 10^{-4}\,M_{\sun}$. The dust emission is located
   within this core.

 \item  We interpret the on-going mass loss as a disk wind, driven by the
   ionization and heating of the gas. The thermal gas velocities are similar
   to the rotation velocity, making the disk susceptible to evaporation. The
   fast wind from the star may be accelerating gas shed from the disk, as seen
   in the [\ion{O}{iii}] line wings. The mass and current mass-loss rate of
   the disk indicate an decay time similar to the age of the nebula.

 \item By comparing masses of dust disks in planetary nebulae of different
   ages, we show that the dust disks dramatically reduce in mass as the nebula
   evolves, reflecting the disk evaporation.  The dust mass decrease from $\sim
   5 \times 10^{-4}\,\rm M_\odot$ at the start of the PN phase, to  $\sim
   4 \times 10^{-7}\,\rm M_\odot$ for the helix nebula on the  white dwarf
   cooling track where
   the luminosity of the star quickly drops by a factor of 100, reducing the
   rate at which the dust disk would be lost.

\end{enumerate}

\begin{acknowledgements}
This work was financially supported 
by MNiSW of Poland through grant No. N\,203\,024\,31/3879. The VLT/UVES
observations are from ESO program 075.D-0104, the VISIR data from  077.D-0652and  the archival data came from
program 077.D-0652.
\end{acknowledgements}

\end{document}